# ATLAS Pixel Opto-Electronics

K.E. Arms, K.K. Gan, P. Jackson, M. Johnson, H. Kagan, R. Kass, A.M. Rahimi,
C. Rush, S. Smith, R. Ter-Antonian, M.M. Zoeller

Department of Physics, The Ohio State University, Columbus, OH 43210, USA

P. Buchholz, M. Holder, A. Roggenbuck, P. Schade, M. Ziolkowski
Fachbereich Physik, Universitaet Siegen, 57068 Siegen, Germany

## ABSTRACT

We have developed two radiation-hard ASICs for optical data transmission in the ATLAS pixel detector at the LHC at CERN: a driver chip for a Vertical Cavity Surface Emitting Laser (VCSEL) diode for 80 Mbit/s data transmission from the detector, and a Bi-Phase Mark decoder chip to recover the control data and 40 MHz clock received optically by a PIN diode. We have successfully implemented both ASICs in 0.25 μm CMOS technology using enclosed layout transistors and guard rings for increased radiation hardness. We present results of the performance of these chips, including irradiation with 24 GeV protons up to 61 Mrad ($2.3 \times 10^{15}$ p/cm$^2$).

# I. INTRODUCTION

The ATLAS pixel detector [1] consists of two barrel layers, two forward, and two backward disks which provide at least two space point measurements. The pixel sensors are read out by front-end electronics controlled by the Module Control Chip (MCC). The low voltage differential signal (LVDS) from the MCC is converted by the VCSEL Driver Chip (VDC) into a single-ended signal appropriate to drive a Vertical Cavity Surface Emitting Laser (VCSEL). The optical signal from the VCSEL is transmitted to the Readout Driver (ROD) via a fiber.

The 40 MHz beam-crossing clock from the ROD, bi-phase mark (BPM) encoded with the command signal to control the pixel detector, is transmitted via a fiber to a PIN diode. This BPM encoded signal is decoded using a Digital Opto-Receiver Integrated Circuit (DORIC). The clock and data signals recovered by the DORIC are in LVDS form for interfacing with the MCC.

The ATLAS pixel optical link system contains 448 VDCs and 360 DORICs with each chip containing four channels. The chips couple to 224 VCSEL and 180 PIN array optical packages (opto-packs). These optical packages are the 8-channel version of the 12-channel optical packages used in the off-detector opto-electronics of the Semi Conductor Tracker (SCT), a silicon strip detector located at the outer radius [2]. The chips and opto-packs are mounted on 180 chip carrier boards (opto-boards).

The optical readout system will be exposed to a large dosage of radiation. We assume that the main radiation effect is surface damage in the CMOS devices due to ionizing radiation and bulk damage in the VCSEL and PIN with the displacement of atoms. We use the Non Ionizing Energy Loss (NIEL) scaling hypothesis to estimate the fluences [3,4,5]. The silicon components (DORIC/VDC and PIN) of the pixel optical link system are expected to be exposed to a maximum total fluence of $3.7 \times 10^{14}$ 1-MeV $n_{eq}/cm^2$ during ten years of operation at the LHC. The corresponding fluence for the GaAs component (VCSEL) is $2 \times 10^{15}$ 1-MeV $n_{eq}/cm^2$. We study the response of the optical link to a high dose of 24 GeV protons. The expected equivalent fluences at LHC are 6.3 and $3.8 \times 10^{14}$ $p/cm^2$, respectively. For simplicity, we present the results from the irradiations with dosage expressed in Mrad using the conversion factor, 1 Mrad = $3.75 \times 10^{13}$ $p/cm^2$.

In this paper we describe the development and testing of the radiation-hard VDC and DORIC circuits for use in the ATLAS pixel detector's optical link.

# II. VDC AND DORIC DESIGN HISTORY

The pixel detector design of the VDC and DORIC takes advantage of the development work for similar circuits [6] used by SCT. Both SCT chips attain radiation-tolerance by using bipolar integrated circuits (AMS 0.8 µm BICMOS) and running with large currents in the transistors at 4 V nominal supply voltage. These chips are therefore not applicable for the higher radiation dosage and lower power budget requirements of the pixel detector.

We originally implemented the VDC and DORIC circuits in radiation-hard DMILL 0.8 µm technology with a nominal supply voltage of 3.2 V. Using three submissions from summer 1999 to May 2001, we developed circuits in this technology which met the electrical specifications. However, an irradiation study of the DMILL circuits in April 2001 with 24 GeV protons at CERN showed severe degradation of circuit performance. We concluded that the DMILL technology did not meet the radiation hardness requirement of the ATLAS pixel detector.

We have therefore migrated the VDC and DORIC designs to the standard deep submicron (0.25 µm) CMOS technology, which has a nominal supply voltage of 2.5 V. By employing enclosed layout transistors and guard rings [7], this technology has been proven to be very radiation hard.

# III. VDC CIRCUIT

The VDC is used to convert an LVDS input signal into a single-ended signal appropriate to drive a VCSEL in a common cathode array. The output current of the VDC should be variable between 0 and 20 mA through an external control current[1], with a standing (dim) current of ~1 mA to improve the switching speed of the VCSEL. The nominal operating current for a VCSEL is 10 mA. The electrical output should have rise and fall times (20–80%) between 0.5 and 2.0 ns; nominally 1.0 ns. In order to minimize the power supply noise on the opto-board, the VDC should also have constant current consumption independent of whether the VCSEL output is in the bright (on) or dim (off) state.

Figure 1 shows a block diagram of the VDC circuit. An LVDS receiver converts the differential input into a single-ended signal. The driver controls the current flow from the positive power supply into the anode of the VCSEL. The VDC circuit is therefore compatible with a common cathode VCSEL array. An externally controlled voltage ($V_{Iset}$) determines the current $I_{set}$ that sets the amplitude of the VCSEL current (bright minus dim current), while another externally controlled voltage (Tunepad) determines the dim current. The driver contains a dummy driver circuit which, in the VCSEL dim state, draws an identical amount of current from the positive power supply as is flowing through the VCSEL in the bright state. This enables the VDC to have constant current consumption.

## IV. DORIC CIRCUIT

The function of the DORIC is to decode BPM encoded clock and data signals received by a PIN diode. Figure 2 shows an example of a BPM encoded signal. The 40 MHz beam-crossing clock is encoded by sending transitions corresponding to clock leading edges only. In the absence of data bits (logic level 0), the resulting signal is a 20 MHz square wave. Data bits are encoded as extra transitions at beam crossing clock trailing edges. The resulting signal for a string of logic level 1 is a 40 MHz square wave. The delay between the decoded data and clock is fixed, simplifying the timing in the MCC chip. In addition, the BPM signal has 50% duty cycle, hence there is no baseline shift due to a different number of data bits being transmitted. The amplitude of the current from the PIN diode is expected to be in the range of 40 to 1000 μA. The 40 MHz clock recovered by the DORIC is required to have a duty cycle of (50 ± 4)% with a total timing error (jitter) of less than 1 ns. The MCC can accept the recovered clock and decoded data LVDS over a wide range of signal average (1.0–1.5 V) and amplitude (0.2–0.5 V). The delay of the data with respect to the clock must be less than 3 ns.

Figure 3 shows a block diagram of the DORIC circuit. In order to keep the PIN bias voltage (up to 15 V) off the DORIC, we employ a single-ended preamp circuit to amplify the current produced by the PIN diode. Since single-ended preamp circuits are sensitive to power supply noise, we utilize two identical preamp channels: a signal channel and a noise cancellation channel. The signal channel receives and amplifies the input signal from the anode of the PIN diode, plus any noise picked up by the circuit. The noise cancellation channel amplifies noise similar to that picked up by the signal channel. This noise is then subtracted from the signal channel in the differential gain stage. To optimise the noise subtraction, the input load of the noise cancellation channel is matched to the input load of the signal channel (PIN capacitance) via a dummy capacitance.

It is important that the preamp and gain stage do not distort the duty cycle of the input signal. Any distortion results in an odd-even jitter in the recovered clock. The gain stage is essentially a comparator, converting the linear output of the preamps to logic signals. It incorporates internal feedback to adjust its threshold so that the 50% duty cycle is maintained over the entire input signal range, 40 to 1000 μA.

In Figure 4 we show a block diagram of the DORIC logic circuitry. A selection of simulated internal logic signals is shown in Figure 5. An edgedet circuit produces a short pulse from each input transition of the gain stage output. Some of these pulses correspond to the leading edge of the 40 MHz system clock while other pulses correspond to data logic level 1 signals. The clock recovery circuitry must be taught how to distinguish between these. This is accomplished after power-up during an initialization period in which only logic level 0 is transmitted. The resulting signal for this case is a 20 MHz square wave. The output of the edgedet is a 40 MHz train of short pulses, one for each edge of the input signal. These are applied to dflip #1, which is connected as a toggle with feedback delayed by delay #1, #2 and #3. The nominal delay of each delay stage is 6.25 ns giving a total delay of 18.75 ns, which is less than the 25 ns period of the 40

[1] The specification for the maximum output current is inherited from the SCT requirement for efficient annealing of the VCSEL.

MHz pulse train. Thus dflip #1 toggles and its output is a 20 MHz square wave. The two inputs to the exclusive "OR" are approximately 90 degrees out of phase. This results in a recovered clock at the exclusive "OR" output with duty cycle correlated to the exact delay of delay #2 and delay #3. The recovered clock is the input to dcontrol, a circuit that converts duty cycle to a DC voltage. This voltage is used to control the delay of delay #2 and #3. Together these blocks, exclusive "OR", delay #2, delay #3 and dcontrol form a delay-lock-loop keeping the recovered duty cycle very close to 50%. This initialisation period must last long enough, about 1ms, for the delay locked loop to settle.

Once the loop is locked, data pulses can be sent. These will be ignored by dflip #1, as can be seen by examining delay #3 and edgedet in Figure 5. For data recovery the edgedet pulses are sent through a short delay to dflip #2 whose "D" input is the recovered clock. The timing guarantees that sending a logic level 1 sets dflip #2. Dflip #3 synchronizes the recovered data to the recovered clock and increases it's width to the full recovered clock period creating a standard NRZ data signal.

A reset line, RstCTRL, allows for slow and controlled recovery of the delay control circuit without the need to cycle the DORIC power.

## V. RESULTS FROM DEEP SUBMICRON SUBMISSIONS

We had five prototype submissions between summer of 2001 and summer of 2003 using three metal layers. Results from the submissions can be found in [9]. Below we summarize the results from the production run. For cost saving purpose, the chips were produced with the MCCs that required five metal layers.

We show the results of various electrical and optical measurements from a sample of eight 4-channel VDCs and DORICs. The chips were mounted in 44-pin packages for measurements on a test board, hence the performance of the chips is somewhat degraded due to parasitic capacitance from the packages. Figure 6 shows the VCSEL current generated by the VDCs as a function of the external control current Iset. The current was measured with a 1-Ω resistor in series with the VCSEL. The dim current is ~1 mA as designed. The bright current saturates above the Iset value of ~0.8 mA. The saturation current is determined by the effective resistance of the VCSEL and maximum voltage provided by the VDC for a 2.5 V supply voltage. The saturation current is ~13 mA in this setup, significantly below the target 20 mA due to the large effective resistance of the VCSEL. Our irradiation studies with protons show that this is adequate to anneal the VCSEL. Furthermore, we have verified that the saturation current is larger with the VCSELs that are used in the production opto-boards. The VDCs have a fairly balanced current consumption. Figures 7 and 8 show the optical rise and fall times of the VCSEL as driven by the VDCs. The fall time is under 1.0 ns. The rise time is typically ~1 ns. A few channels have rise times as large as 1.5 ns. The rise time as measured on the opto-board is under 1.0 ns for all channels over the operating range of the circuit due to the more optimized board layout.

All the measurements described below for the DORICs were performed at the PIN current of 100 μA. Measurements at 40 and 800 μA yield similar results. The current consumption of the DORIC is ~75 mA for 4 channels at the supply voltage of 2.5 V. The amplitude and average of both clock and command LVDS are within the specification. The duty cycle of the decoded 40 MHz clock is within the specification of (50 ± 4)%. The clock jitter is also within the specification, < 1.0 ns, as shown in Figure 9. The rise and fall times of both clock and command are under 1 ns and the delay of the command with respect to the clock is within the specification. DORIC thresholds for no bit errors as measured on the test board are typically under ~65 μA. The thresholds as measured on the opto-board are lower, < 20 μA, due to the more optimized board layout as noted above. Since the thresholds are low, no capacitors are attached to the dummy input channels for noise cancellation.

## VI. OPTO-BOARD

The VDCs, DORICs, and opto-packs are mounted on opto-boards. We produce opto-boards of two flavours with each opto-board serving six or seven modules. To maximize the number of spare opto-boards, all opto-boards are fabricated to serve seven modules. For the outer barrel layer and the disk system each module requires one link for transmitting data, therefore each opto-board contains two VDCs plus one VCSEL array opto-pack on the top side and two DORICs plus one PIN array opto-pack on the bottom side. The inner barrel (B) layer is expected to have a higher hit occupancy, therefore each module requires two links for transmitting data to the ROD. The top side hence contains four VDCs plus two VCSEL array opto-packs with each pair of links serving a module connected to adjacent fibers in an 8-fiber ribbon.

The layout of the opto-boards is shown in Figure 10. The bottom layout is the same for both opto-board flavours. The supply voltages are connected via the 80-pin connector on the bottom side. The input and output LVDS are also fed through the connector. The opto-board is fabricated with beryllium oxide (BeO) as the substrate for heat management. On the bottom side, no component is placed between the connector and the mounting hole near the edge of the board to allow for thermal contact with the cooling pipe.

We initially prototyped the opto-boards using FR-4 as the substrate for fast turnaround and cost saving. Four prototype runs were submitted to accommodate changes in the chips and opto-pack design. We had two prototype submissions with BeO. The first vendor produced opto-boards with many open vias and shorts between traces. However, the second vendor [10] was able to produce opto-boards of high quality. We observe low PIN current thresholds for no bit errors when all links are active, indicating that there is no significant cross talk on the opto-board.

## VII. IRRADIATION STUDIES

We have irradiated VDCs and DORICs from three deep submicron submissions with 24 GeV protons at CERN. We use two setups during each irradiation. In the first setup ("cold box"), we perform electrical testing of VDCs and DORICs at a controlled temperature of -10 $^{o}$C. No optical components (VCSEL or PIN) are used, allowing direct study of the possible degradation of the chips without additional complications from the optical components. For the VDCs, we monitor the rise and fall times and the bright and dim currents. For the DORICs, we monitor the minimum input current for no bit errors, clock jitter and duty cycle, rise and fall times and the amplitude and average of the recovered clock and command LVDS.

In the second setup ("shuttle"), we test the performance of the opto-link using opto-boards. The opto-boards mounted in this setup may be remotely moved in and out of the beam target area. In the control room, we generate bi-phase mark encoded pseudo-random signals for transmission via 25 m of optical fibers to the opto-boards. The PIN diodes on the opto-boards convert the optical signals into electrical signals. The DORICs then decode the electrical signals to extract the clock and command LVDS. The LVDS are fed into the VDCs and converted into signals that are appropriate to drive the VCSELs. The optical signals are then sent back to the control room for comparison with the generated signals. We remotely move the opto-boards on the shuttle out of the beam to anneal the VCSELs. We typically irradiate the opto-boards for 10 hours (~5 Mrad), then anneal the VCSELs for the rest of the day with large current (~14 mA or higher in the last irradiation depending on the effective resistance of the VCSEL).

We report the result from the last irradiation using VDCs and DORICs from the production run.

In the cold box, we irradiated four DORICs and four VDCs in summer of 2003. We observe no significant degradation in the chip performance up to a total dose of ~61 Mrad, except the average of the clock LVDS of one DORIC, which increases by 10%, an acceptable change. Some examples of comparisons of the performance before and after irradiation are shown below. Figure 11 shows a comparison of the bright and dim currents of the VDCs. Figures 12 and 13 show a comparison of the rise and fall times of the decoded clock of the DORICs.

In the shuttle setup, we irradiated four opto-boards with the dosage of (8-12) x $10^{14}$ p/cm$^2$ (22-32 Mrad) in summer of 2004. The PIN current thresholds for no bit errors, as observed for over 10 seconds (4 x $10^8$ bits), are all below 40 μA and remain constant or decrease slightly up to the total dosage of 25 Mrad, as shown for one opto-board in Figure 14. We also measured bit errors as a function of PIN current during spills to calculate the single event upset (SEU) cross section as shown in Figure 15. Comparing our results with the results from a similar study for the SCT [11], we find that the SEU for both systems are similar, suggesting that the dominant source of SEU effects is energy deposition in PIN diode. Using the estimated particle flux at the location of the opto-board on the pixel detector ($2 \times 10^{6}$ $s^{-1}cm^{-2}$) and SEU cross section, we calculate the bit error rate (BER) as a function of PIN current. The BER decreases with PIN current as expected, as shown in Figure 16 for one of the opto-boards. The BER at 100 μA is ~$3 \times 10^{-10}$. Since the BER for the DORIC is < $10^{-11}$, the opto-link BER is limited by SEU.

The optical power from the opto-boards were monitored in the control room during the irradiation. Figure 17 shows the optical power as a function of time (dosage) for one opto-board. We observe a general trend in the data: during the irradiation the optical power decreases; the optical power increases during the annealing, as expected. The power loss is due to radiation damage to the VCSELs as the VDCs and DORICs show no radiation damage with up to ~61 Mrad of irradiation. The power in all channels are significantly above 350 μW, the specification for absolute minimum power after irradiation. We received the opto-boards from CERN two months after irradiation and have measured their electrical and optical properties. After ~160 hours of annealing at the VCSEL current of ~14 mA, the optical power of all channels were above 1000 μW, hence well above the minimum requirement.

The optical rise and fall times satisfy the specification, < 1.0 ns, as shown in Figure 18. For the clock and command LVDS, the amplitude and average are within the acceptable range and the rise and fall times are below 1.0 ns as shown in Figures 19, 20 and 21. There is no significant degradation in clock duty cycle. A few channels are slightly above the desired range of (50 ± 4)% as shown in Figure 22. The clock jitter is under 1.0 ns as shown in Figure 23.

In summary, the irradiation results show that the VDC and DORIC have no significant degradation up to a dose of 61 Mrad, hence are adequate for the pixel detector opto-link.

## VIII. SUMMARY

We have developed VDC and DORIC circuits in the deep submicron (0.25 μm) technology using enclosed layout transistors and guard rings for improved radiation hardness. The circuits meet all the requirements for operation in the ATLAS pixel optical link and appear to be sufficiently radiation hard for ten years of operation at the LHC.

## ACKNOWLEDGEMENTS

The authors would like to thank D.J. White for discussions on the DORIC and VDC architectures, K. Einsweiler for his advice and Academia Sinica for providing the opto-packs used in the tests. The authors are indebted to Maurice Glaser and Petr Sicho for their help with the irradiation studies at CERN. This work was supported in part by the U.S. Department of Energy under contract No. DE-FG-02-91ER-40690 and by the German Federal Minister for Research and Technology (BMBF) under contract 056Si74.

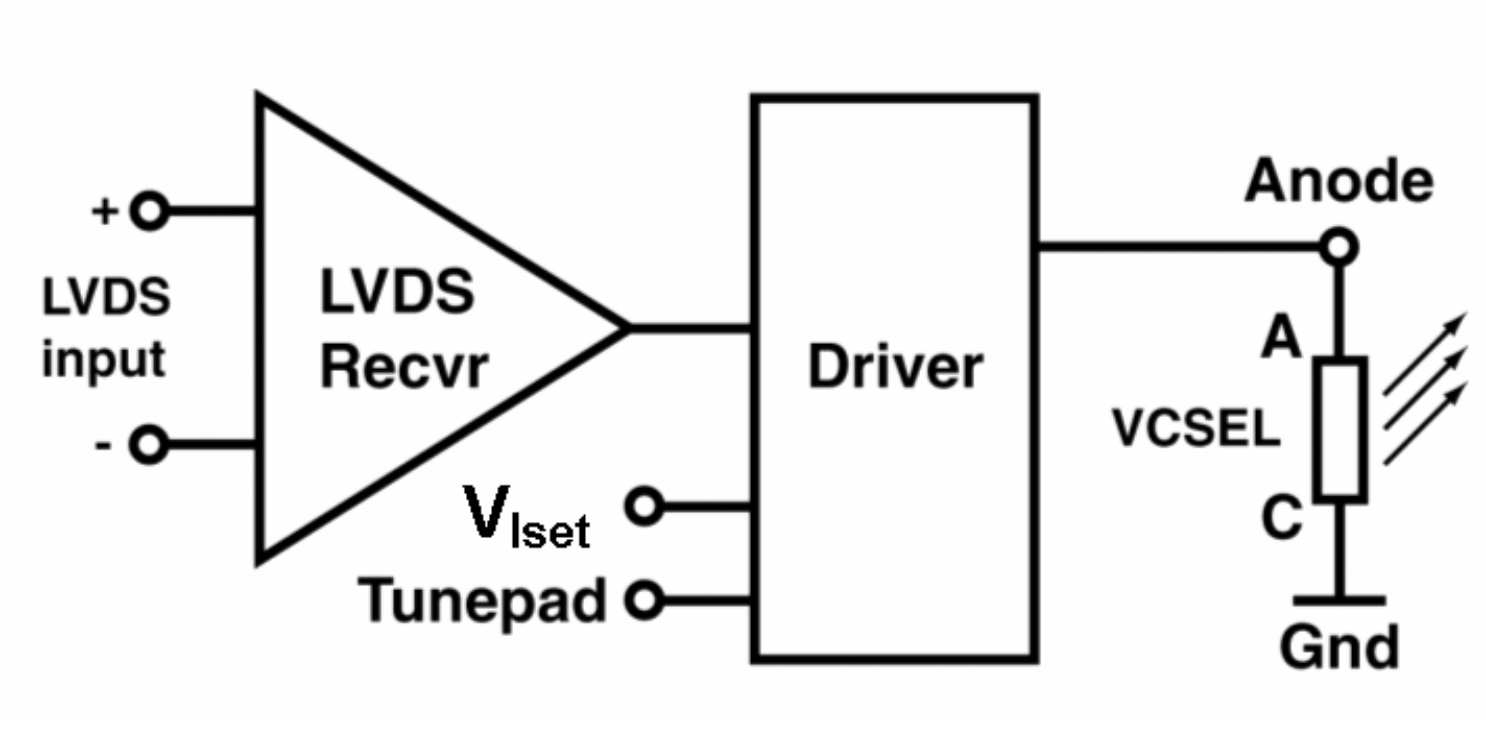

**Figure 1:** Block diagram of the VDC circuit.

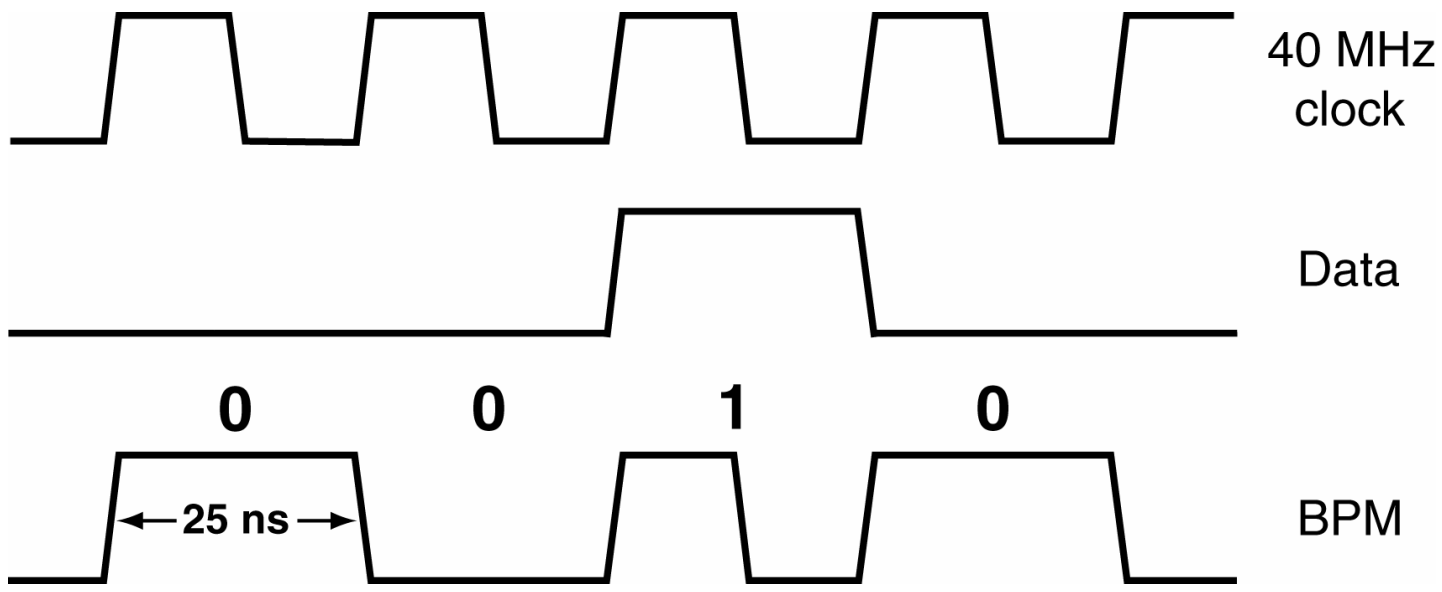

**Figure 2:** Example of a BPM encoded signal.

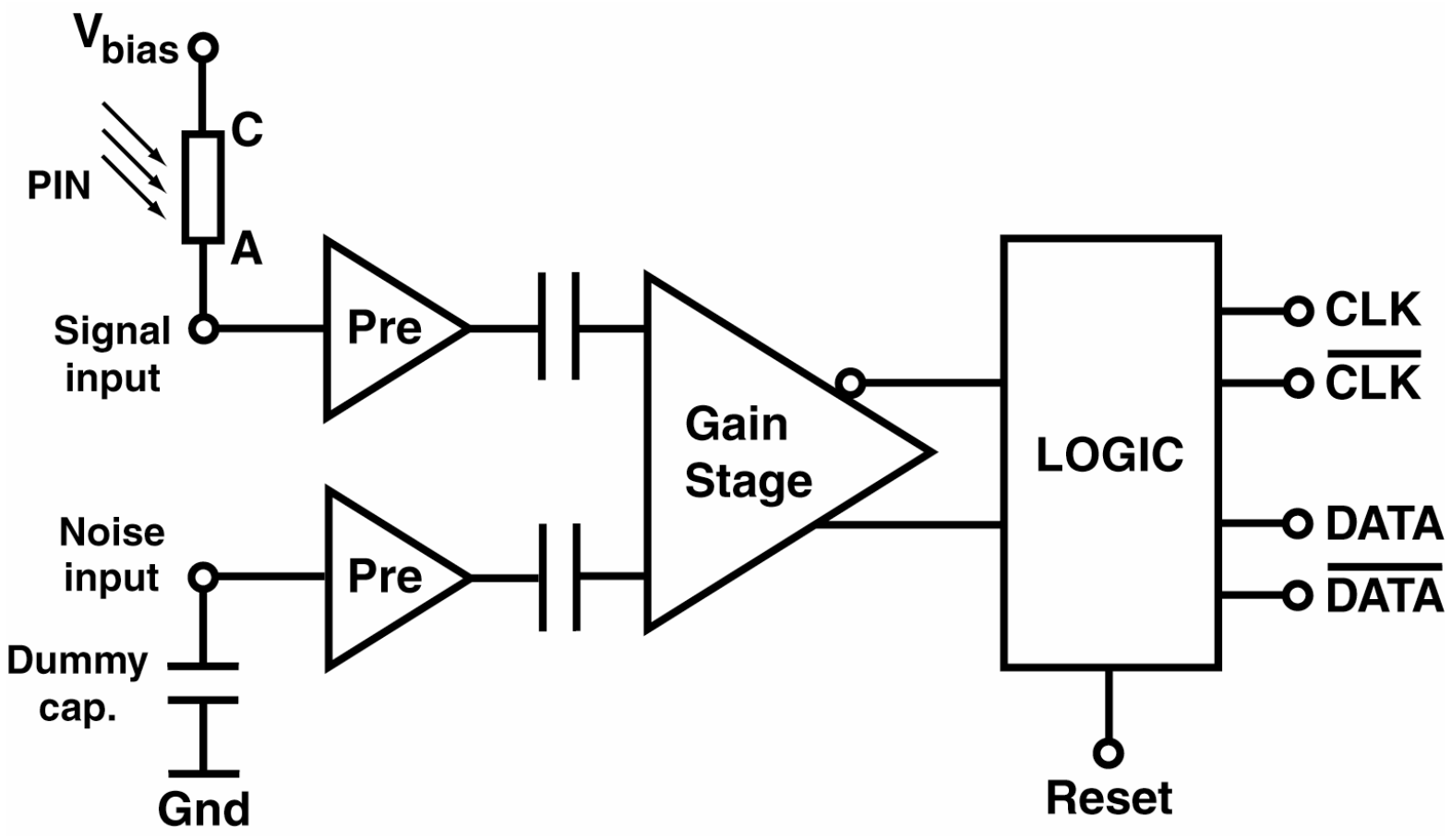

**Figure 3:** Diagram of the DORIC circuit.

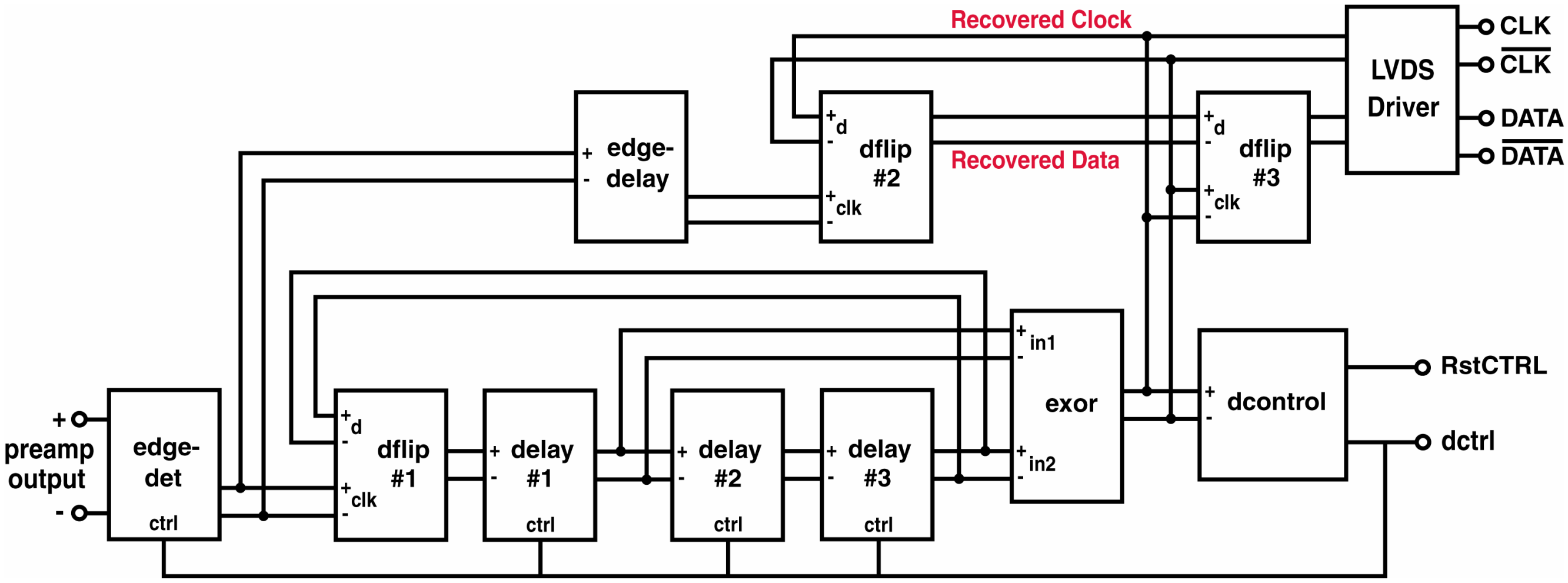

**Figure 4:** Diagram of DORIC logic circuitry.

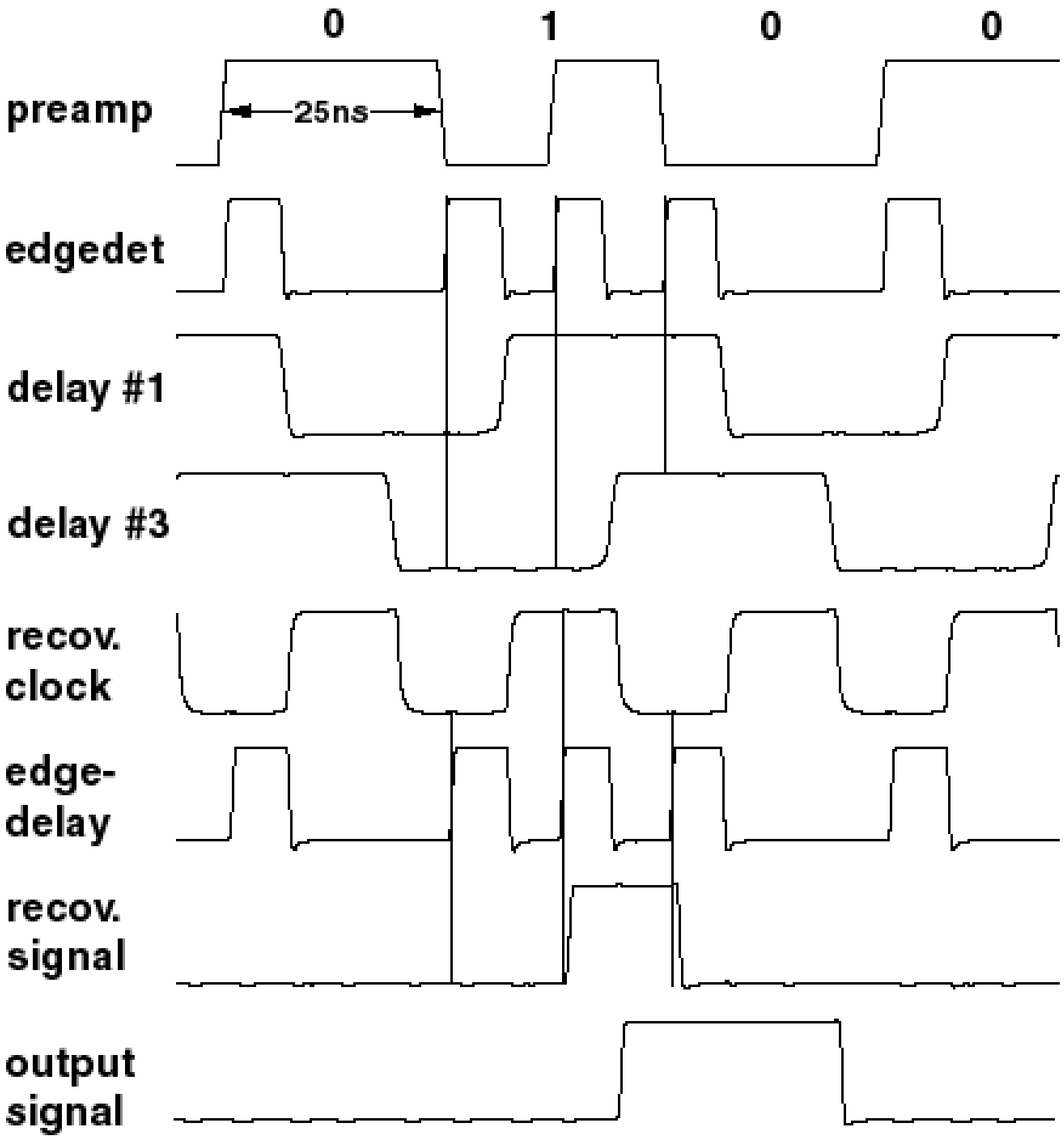

**Figure 5:** Logic signals of DORIC from HSPICE [8] simulation.

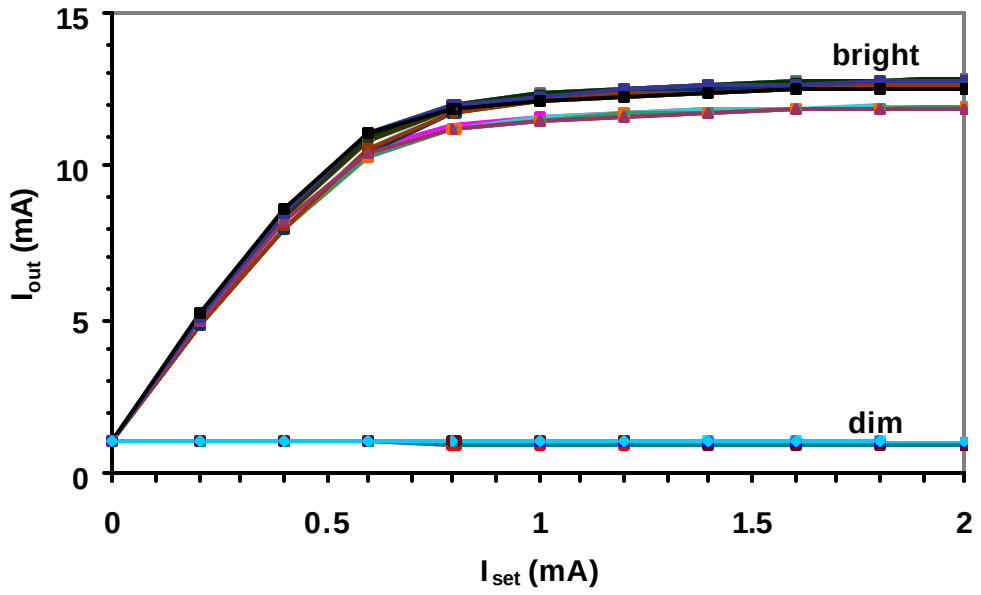

**Figure 6:** VCSEL drive current vs. $I_{set}$ of eight 4-channel VDCs.

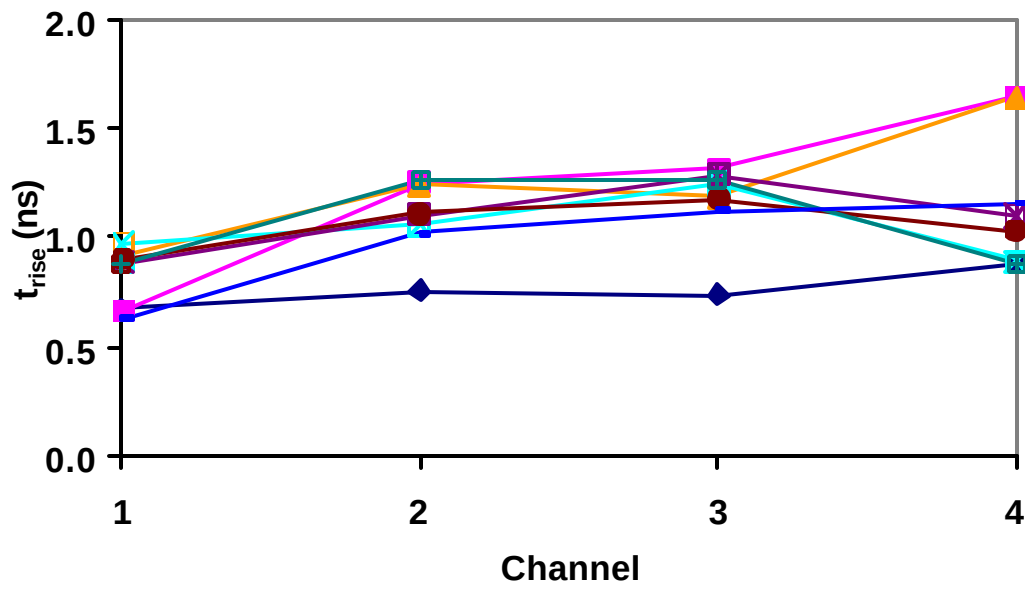

**Figure 7:** Optical rise times of the VCSEL as driven by eight 4-channel VDCs.

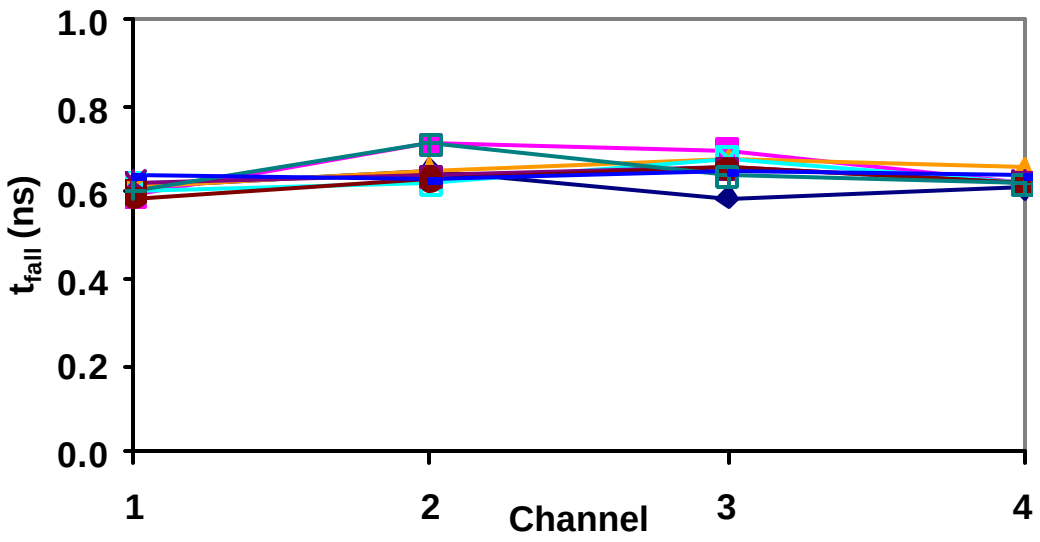

**Figure 8:** Optical fall times of the VCSEL as driven by eight 4-channel VDCs.

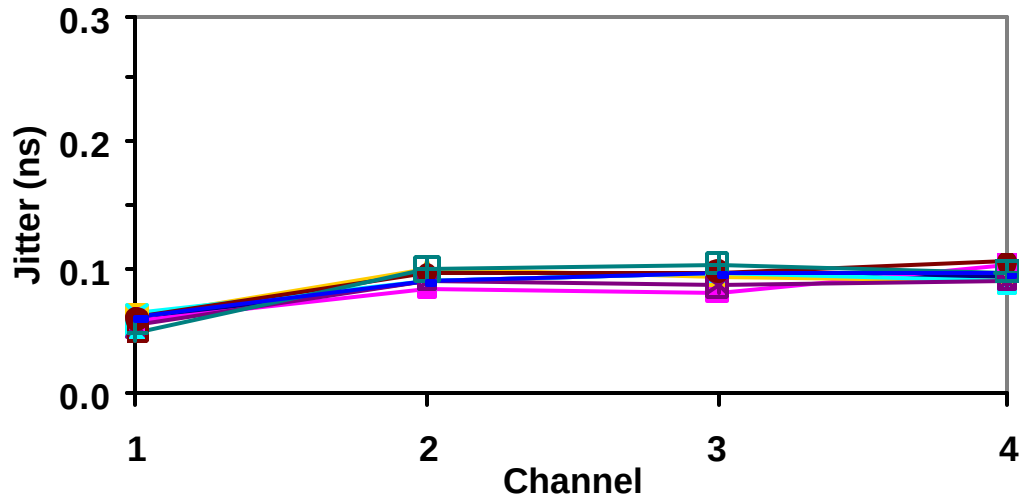

**Figure 9:** Jitter of the decoded clock of eight 4-channel DORICs.

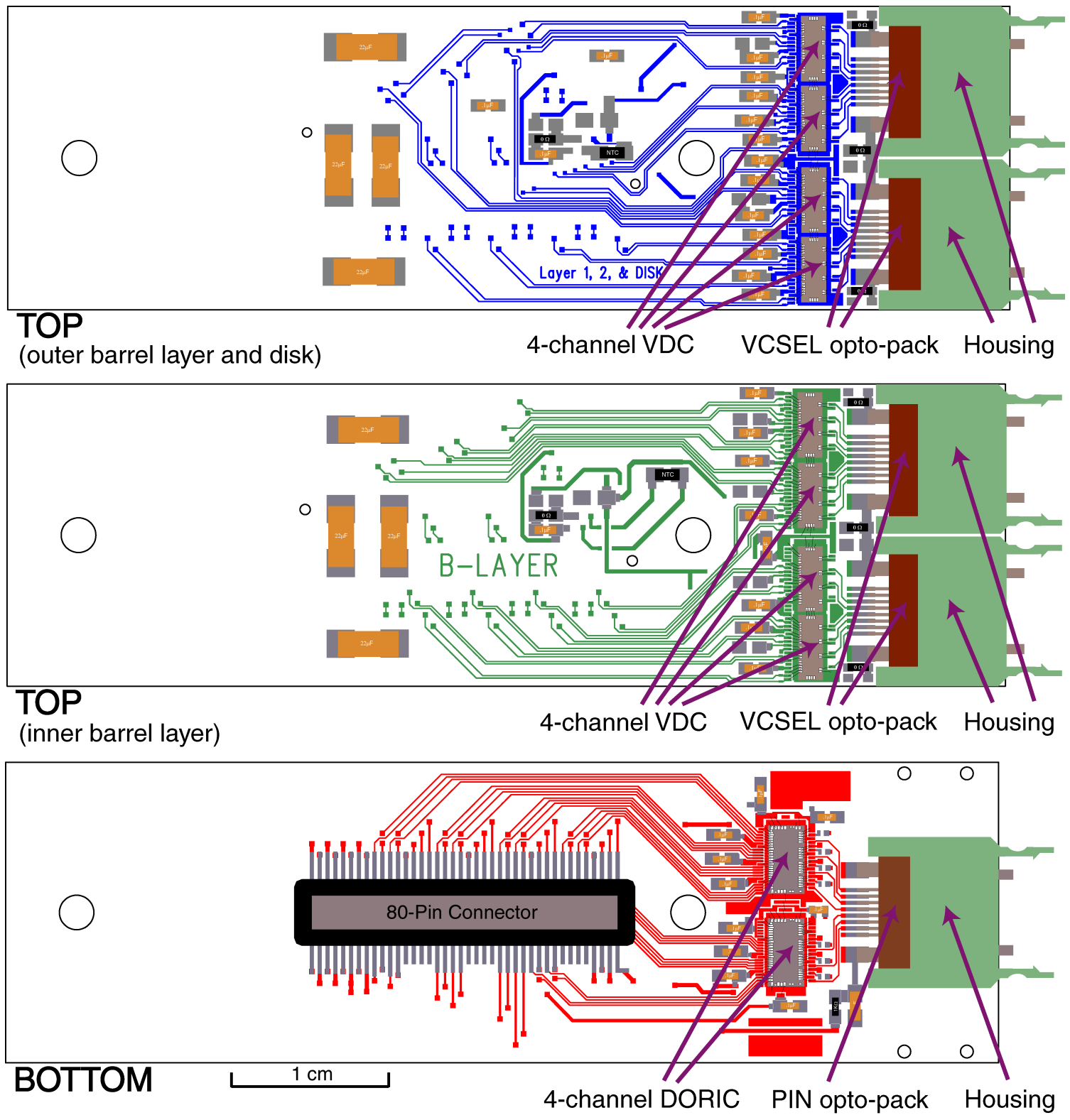

**Figure 10:** The layout of the opto-boards. For outer barrel layer and disk, the top side will be instrumented with only one VCSEL opto-pack and two VDCs to provide the seven links in order to satisfy the mechanical constraints of the opto-board patch panel.

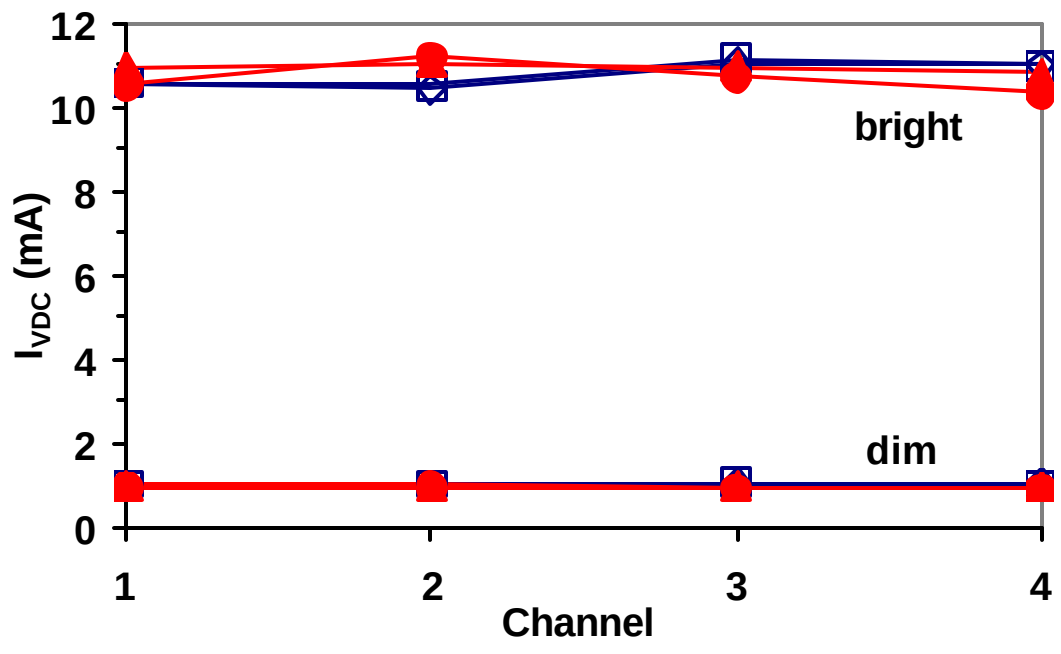

**Figure 11:** VCSEL drive currents of two 4-channel VDCs before (open) and after (filled) irradiation.

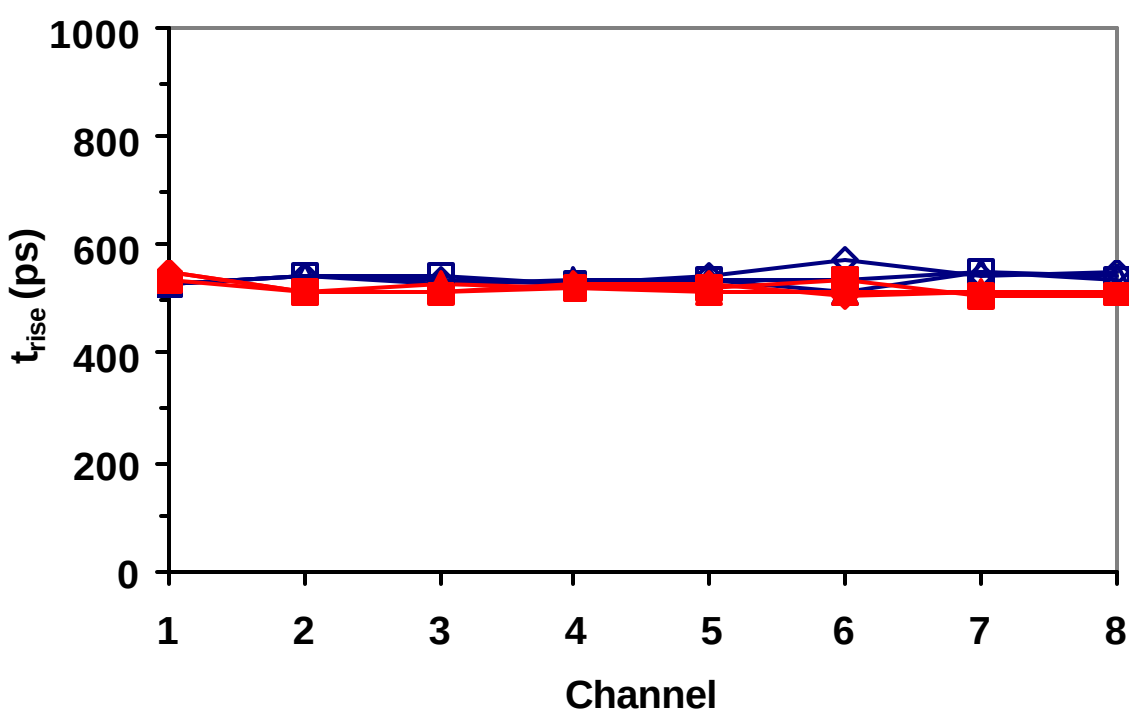

**Figure 12:** Rise times of the decoded clock of two 4-channel DORICs before (open) and after (filled) irradiation.

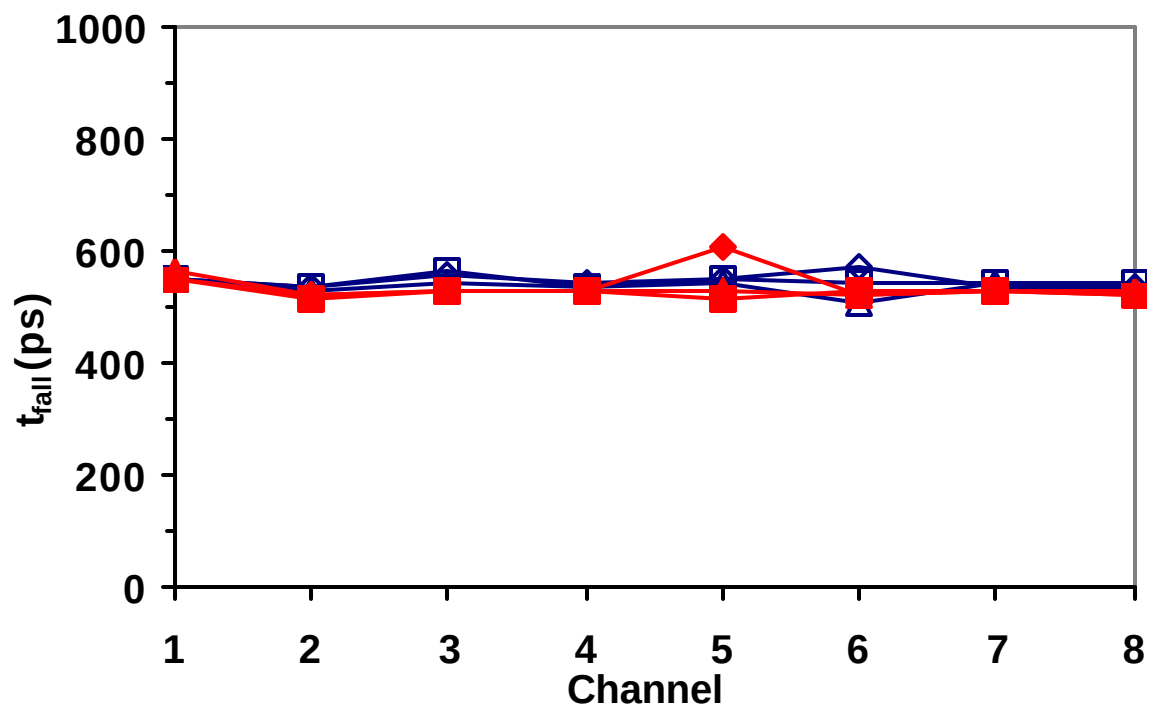

**Figure 13:** Fall times of the decoded clock of two 4-channel DORICs before (open) and after (filled) irradiation.

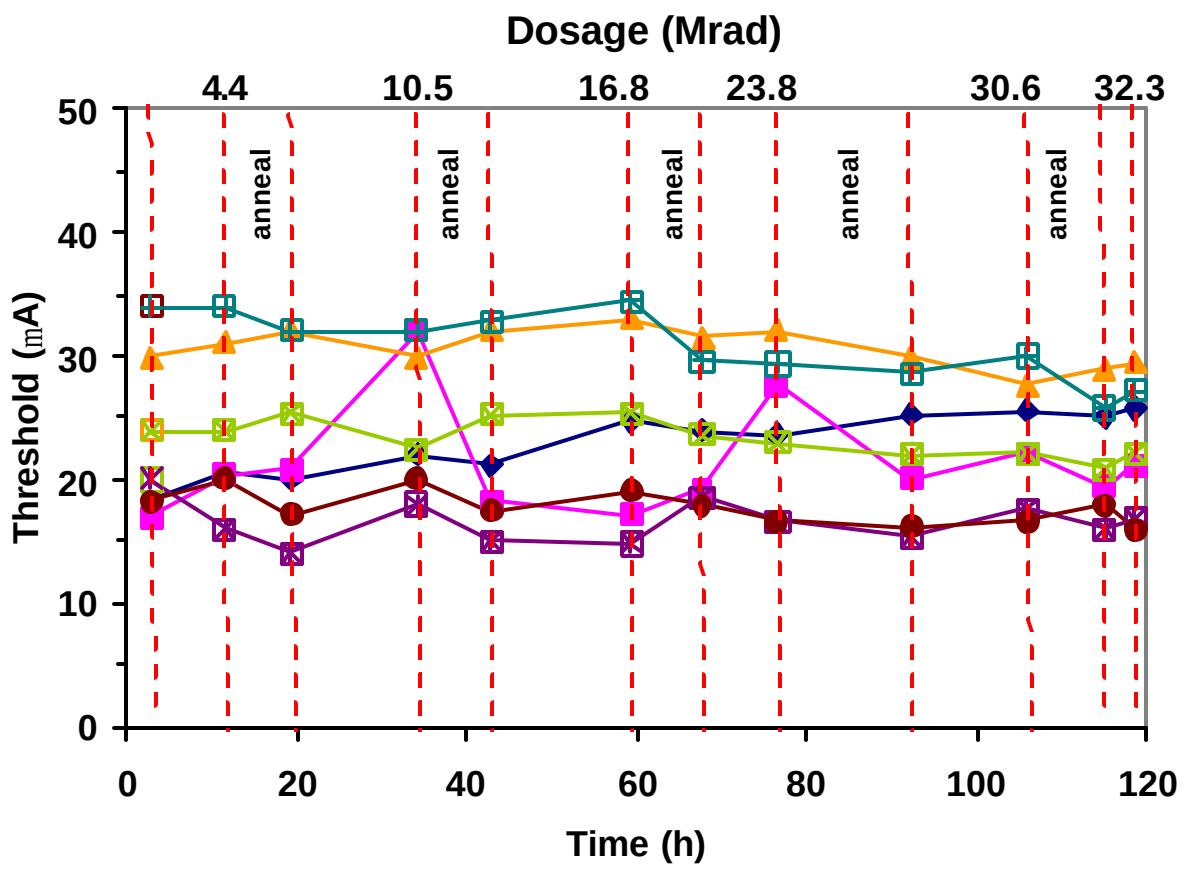

**Figure 14:** PIN current thresholds for no bit errors as a function of time (dosage) for one of the opto-boards with seven active links in the shuttle setup.

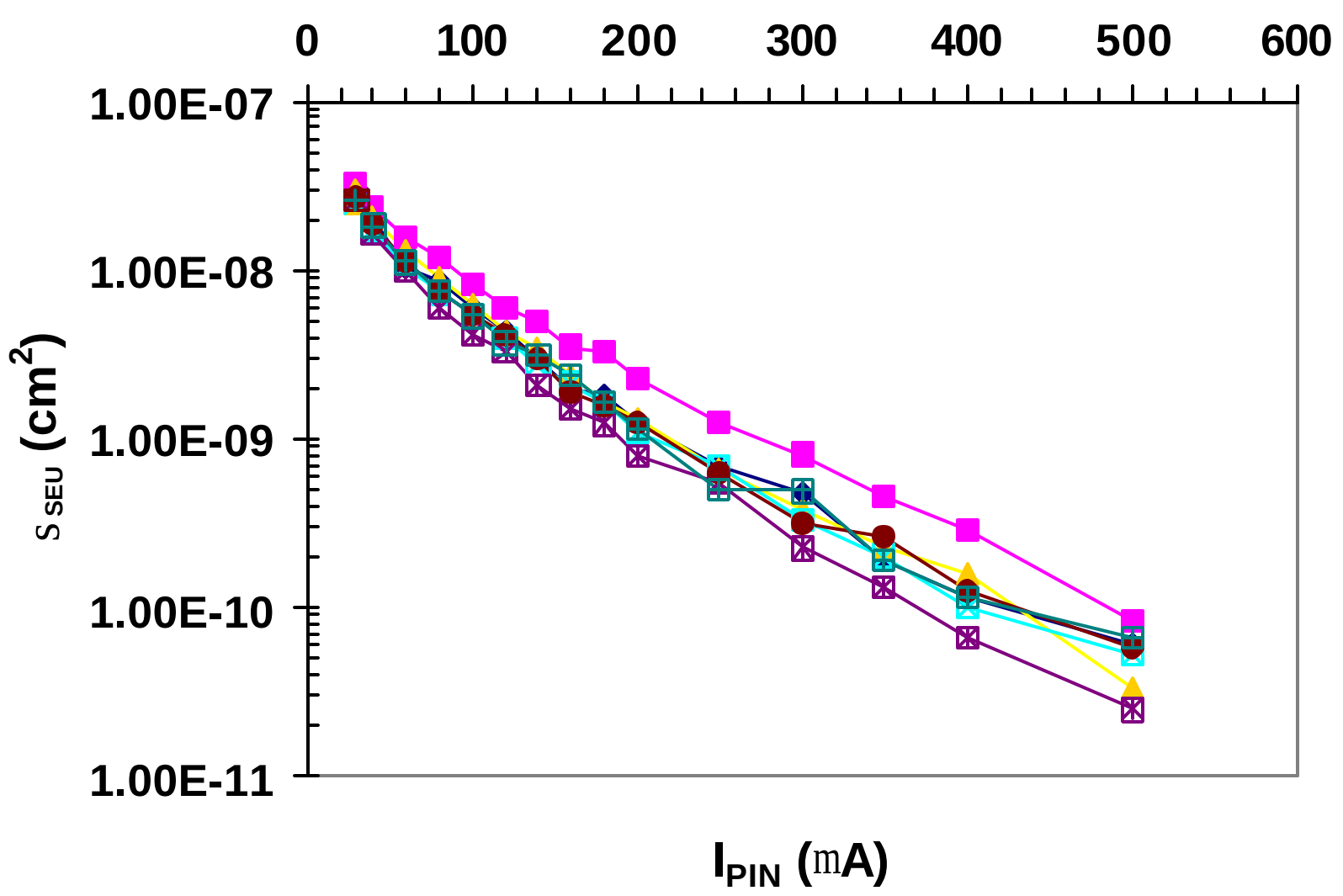

**Figure 15:** Single event upset (SEU) cross section as a function of PIN current for one of the opto-boards in the shuttle setup.

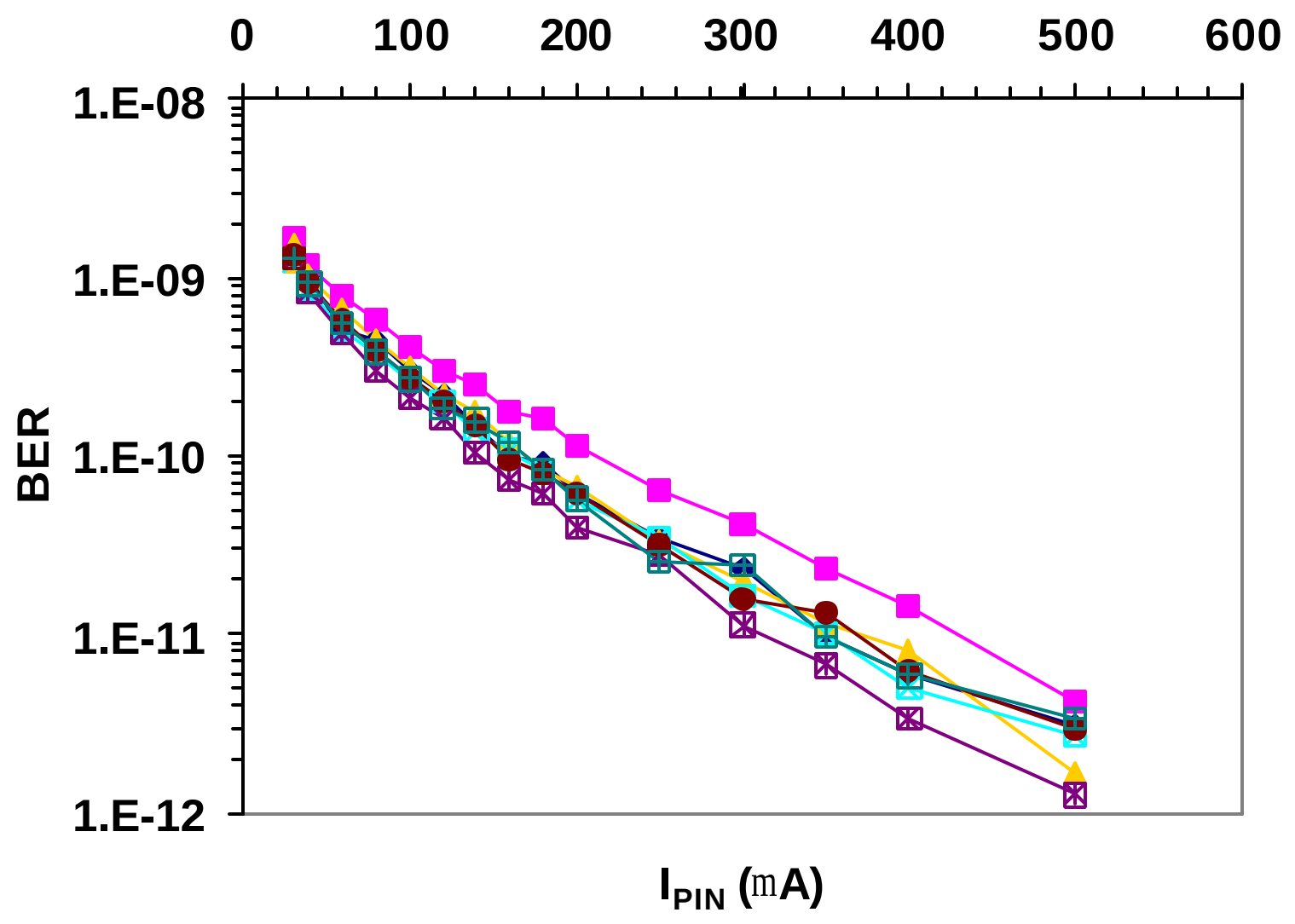

**Figure 16:** Bit Error Rate due to SEU as function of PIN current for one of the opto-boards in the shuttle setup

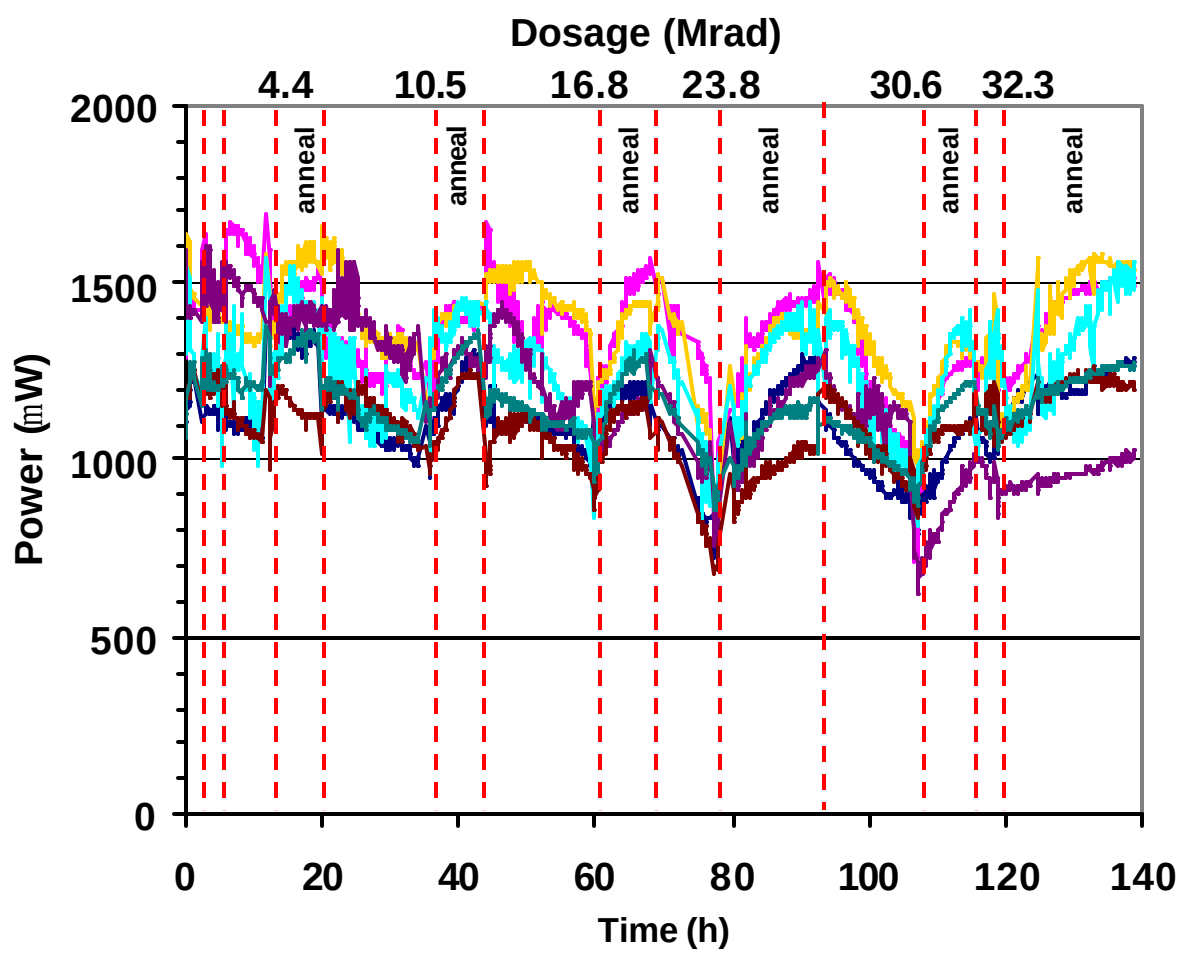

**Figure 17:** Optical power as a function of time (dosage) in the data channels for one of the opto-boards with seven active links in the shuttle setup.

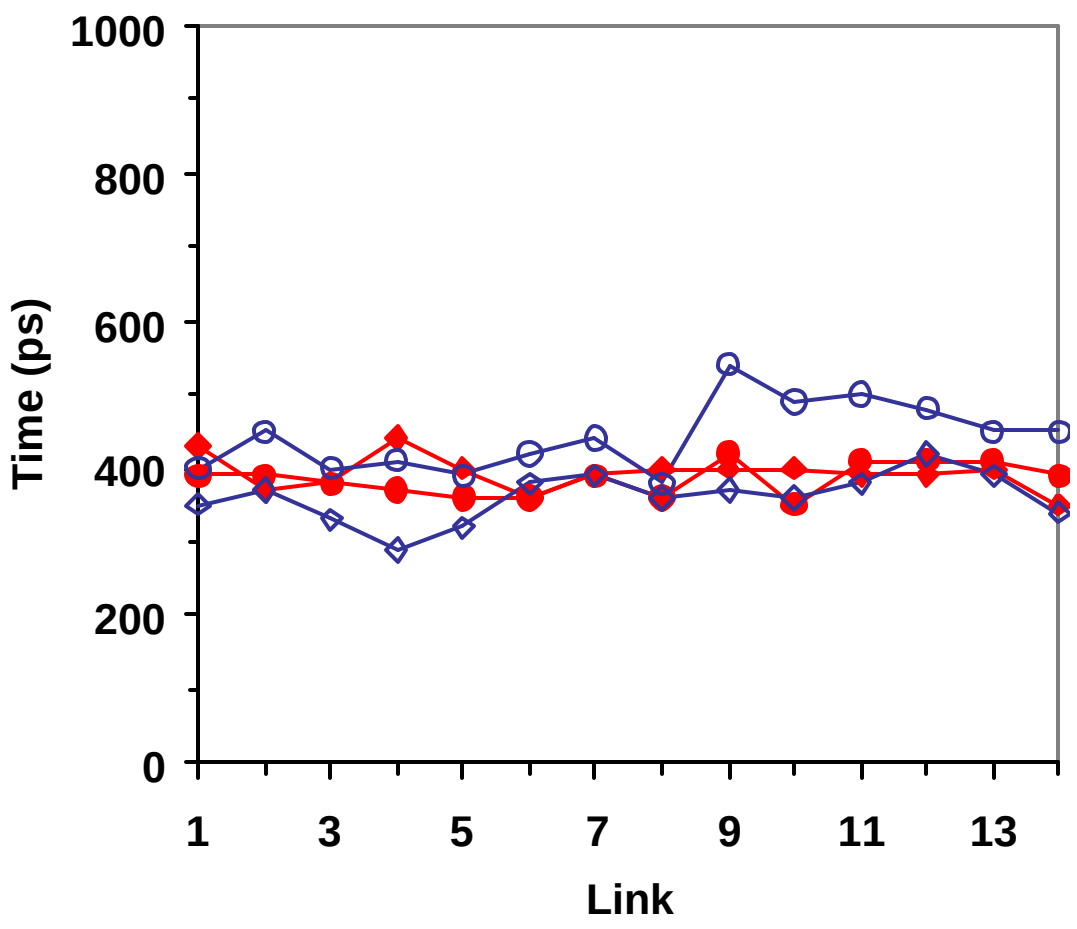

**Figure 18:** Optical rise (diamond) and fall (circle) times of the VCSELs as driven by the VDCs on one of the opto-boards in the shuttle setup before (open) and after (filled) irradiation.

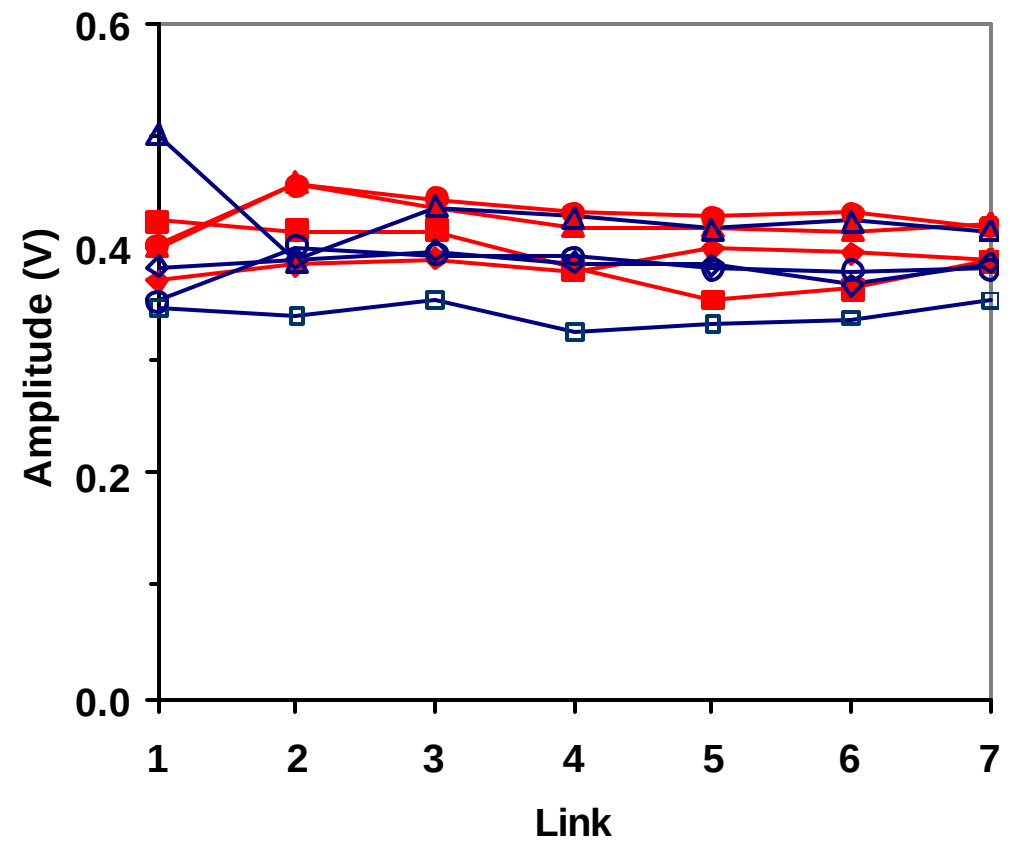

**Figure 19:** Amplitude of clock and data LVDS for two opto-boards in the shuttle setup before (open) and after (filled) irradiation. Both positive and negative polarity of the signals are shown.

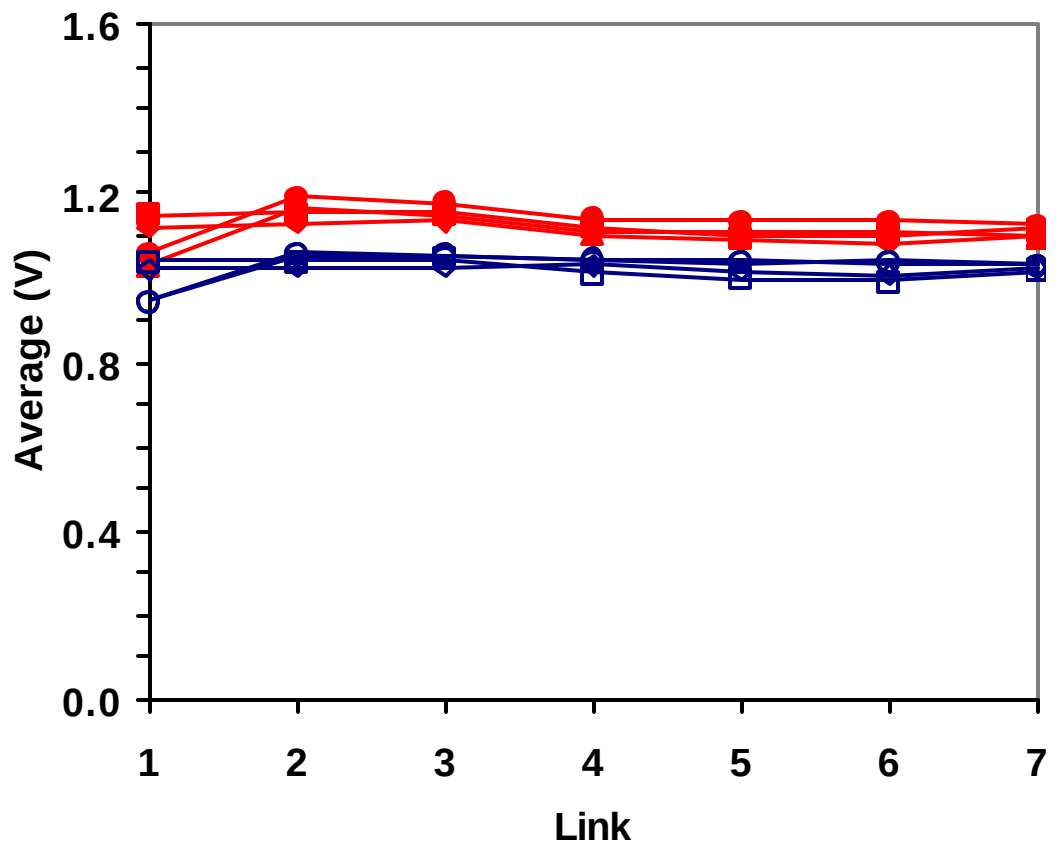

**Figure 20:** Average of clock and data LVDS for two opto-boards in the shuttle setup before (open) and after (filled) irradiation. Both positive and negative polarity of the signals are shown.

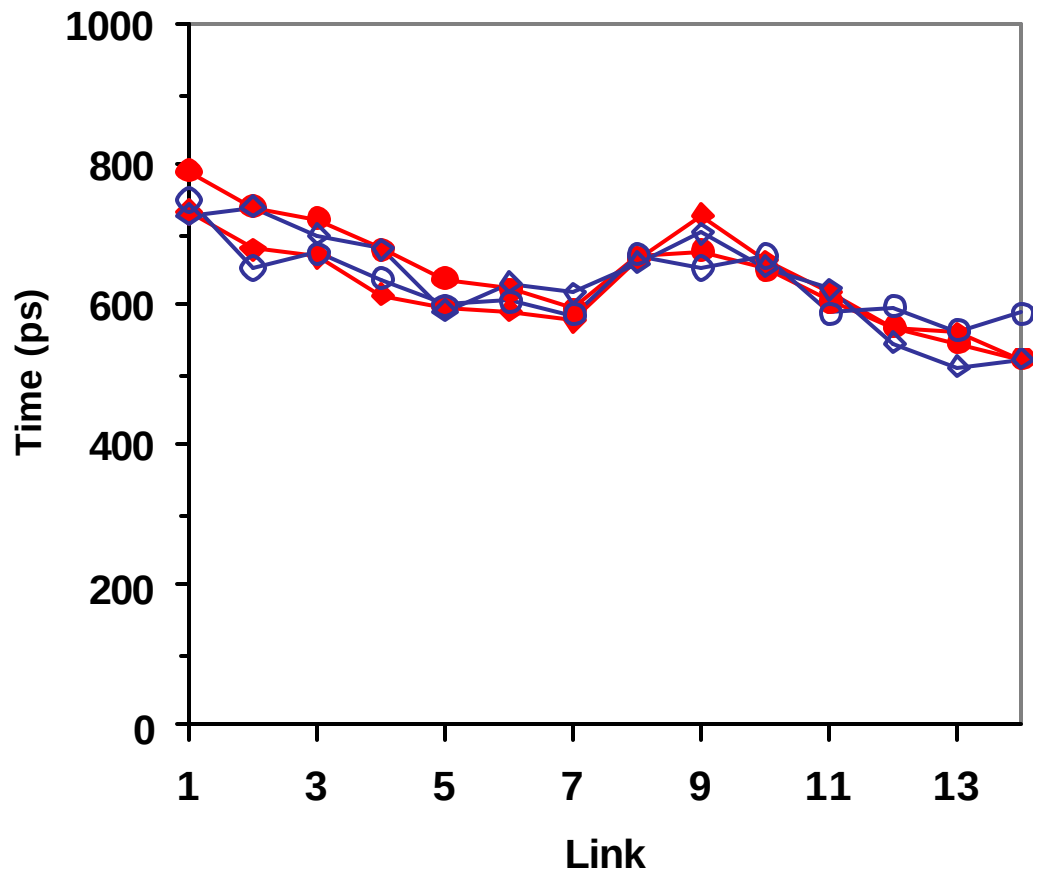

**Figure 21:** Rise and fall times of the LVDS in the shuttle setup before (open) and after (filled) irradiation.

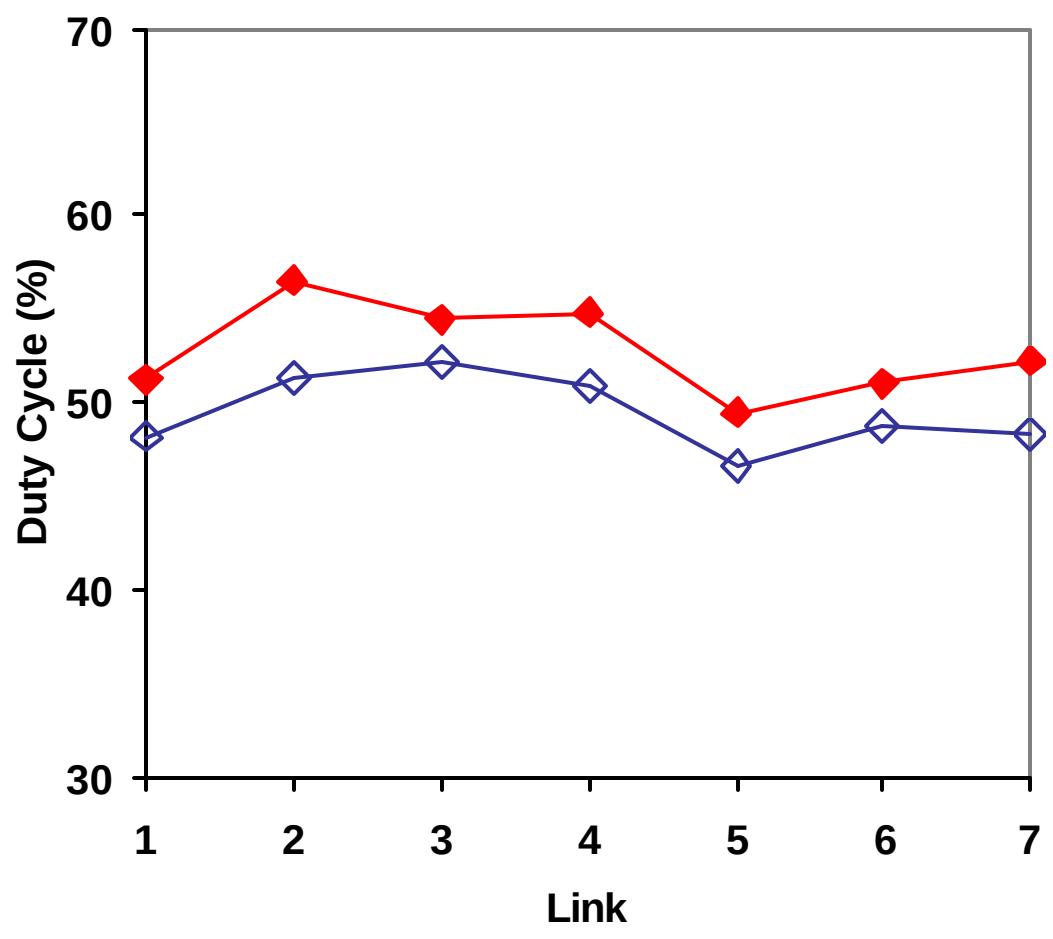

**Figure 22:** Duty cycle of the decoded clock in the shuttle setup before (open) and after (filled) irradiation.

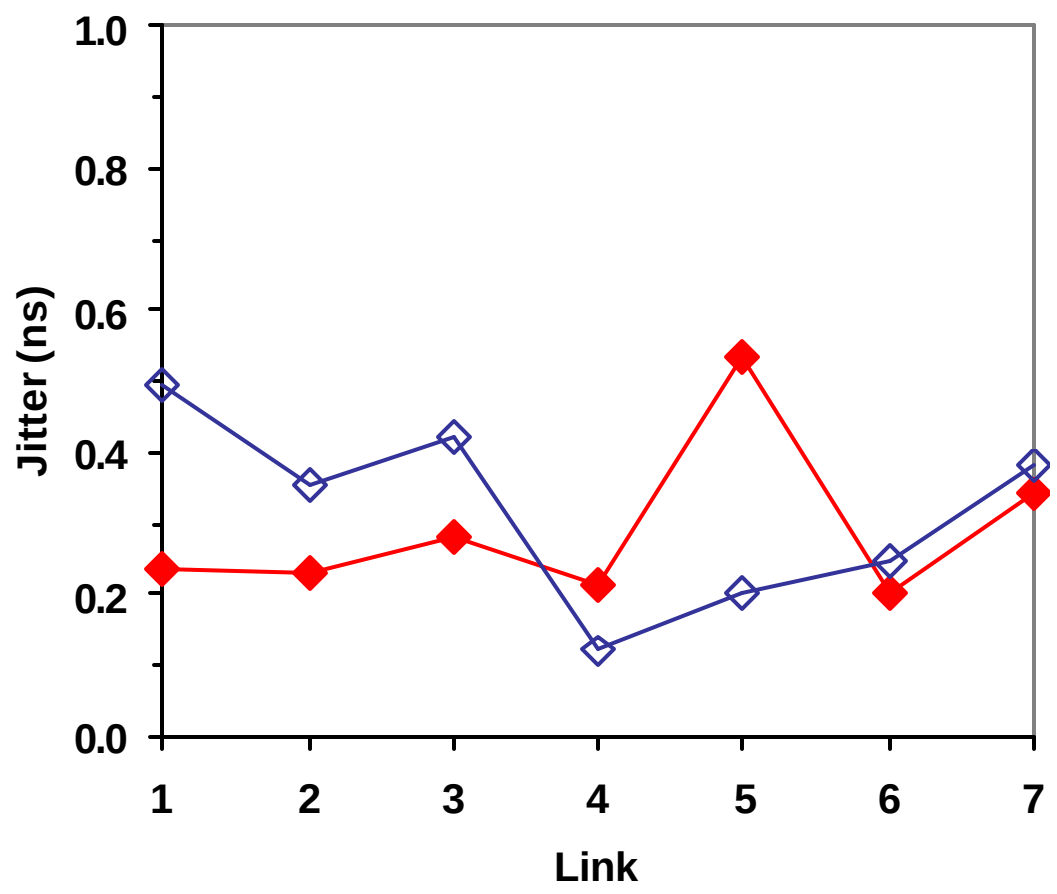

**Figure 23:** Jitter of the decoded clock in the shuttle setup before (open) and after (filled) irradiation.